\documentclass[%
 aps,
 prc,
 reprint,
 superscriptaddress,
 showpacs
]{revtex4-1}

\usepackage{amsmath}
\usepackage{amssymb}
\usepackage{graphicx}% Include figure files
\usepackage{dcolumn}% Align table columns on decimal point
\usepackage{bm}% bold math
\begin{document}

%\preprint{}

\title{Formation of Heavy Meson Bound States by Two Nucleon Pick-up Reactions}



\author{N. Ikeno}
\affiliation{
Department of Physics, Nara Woman's University, Nara 630-8506, Japan}

\author{J. Yamagata-Sekihara}
\affiliation{
Departamento de F$\acute{i}$sica Te$\acute{o}$rica and
IFIC, Centro Mixto Universidad de Valencia-CSIC, Institutos de
Investigaci$\acute{o}$n de Paterna, Aptdo. 22085, 46071 Valencia, Spain}
\affiliation{
Yukawa Institute for Theoretical Physics, Kyoto University, Kyoto
606-8502, Japan }

\author{H. Nagahiro}
\affiliation{
Department of Physics, Nara Woman's University, Nara 630-8506, Japan}

\author{D. Jido}
\affiliation{
Yukawa Institute for Theoretical Physics, Kyoto University, Kyoto
606-8502, Japan }

\author{S. Hirenzaki}
\affiliation{
Department of Physics, Nara Woman's University, Nara 630-8506, Japan}


\date{\today}

\begin{abstract}
We develop a model to evaluate the formation rate of the heavy mesic nuclei
in the two nucleon pick-up reactions, and apply it to the
$^6$Li target cases for the formation of 
heavy meson$-\alpha$ bound states, as examples.
The existence of the quasi-deuteron in the target nucleus is assumed in
this model.
It is found that the mesic nuclei formation in the recoilless kinematics
is possible even for heavier mesons than nucleon in the two nucleon
pick-up reactions.
We find the formation rate of the meson$-\alpha$ bound states can be 
around half of the elementary cross sections at the recoilless
kinematics with small distortions.
\end{abstract}

\pacs{21.85.+d, 36.10.Gv, 25.10.+s}
\keywords{}

\maketitle

%\section{}

\section{Introduction \label{Intro}}
Meson-Nucleus systems are one of the most interesting laboratories to
study the meson properties at finite density and to explore the
symmetry breaking pattern of QCD and its partial restoration in
nucleus~\cite{QCD,Jido,Kolo}.
Especially, the studies of the bound states of meson and nucleus have
the following advantages; 
(i) the selective observation of the meson properties is possible by
making use of the fixed quantum numbers of the bound states, 
(ii) the meson properties inside the nucleus can be observed clearly 
only with relatively small contamination from the vacuum processes, and 
(iii) the system is quasi-static and 
the time dependent dynamical evolution of the system is irrelevant.
These features are different from other methods based on the scattering
and collision processes. 
On the other hand, the
information obtained from the observation of the bound states are
limited in $\rho \lesssim \rho_{0}$ and $T=0$ region in the QCD phase diagram.

We have studied so far the physical interests, the structures, and the
formation reactions of the various kinds of the meson-nucleus
systems~\cite{d3He-nd,K,eta}.
Within these studies, the most exciting and successful results were
obtained by the observation of deeply bound pionic atoms in the one
nucleon pick-up ($d,^3$He) reactions~\cite{Jido,d3He-nd,K.Suzuki}.
We have also considered other one nucleon pick-up reactions like
($\gamma,p$)~\cite{gammap} and ($\pi,N$)~\cite{piN} for the mesic nuclei
formations.
The one nucleon pick-up reactions are found to be useful for the mesic
nucleus formation, however, they require the large momentum transfer for
heavy meson production which is one of the main obstacle to observe
bound states.
Thus, we think that we need to develop new methods for bound state
formations to extend our studies to other meson-nucleus
systems, especially for heavier mesons.
Actually, we are interested in the heavy meson-nucleus systems such as
$\eta^{\prime}(958)$ meson for the studies of $U_{A}(1)$ anomaly
effects~\cite{Nagahiro,Nagahiro2,Nagahiro3}, $\phi$
meson for $\bar{s}s$ components of nucleon and OZI rule at finite
density~\cite{Yamagata}, and $D$
mesons for the charm meson properties in nucleus~\cite{Garcia}.
Thus, we would like to study the two nucleon pick-up reactions
theoretically as a possible method suited to form the heavy meson nucleus
bound systems.

One of the most serious problems in the formation reactions of the heavy
meson nucleus 
systems are the large momentum transfer as mentioned above.
It is known that the matching condition of momentum and angular momentum
transfers plays an important role to determine
the largely populated subcomponents and it is also known that the best
choice for our purpose is the total angular momentum
transfer $J=0$ state formation in the recoilless kinematics in
many cases.
In the one nucleon pick-up reactions which we have mainly considered so far,
the recoilless kinematics can not be satisfied for the formation of
heavier meson bound states than the nucleon mass because of the large
mass of the meson.
Thus, we consider the two nucleon pick-up reactions in this article to
investigate
the possibility to extend our study to heavier meson region using these
reactions.
Actually, there was an attempt to observe the $\eta$-mesic state in
the two nucleon pick-up 
$^{27}$Al($p$,$^{3}$He) reaction at COSY-GEM~\cite{Budzanowski}.

\section{Effective number formalism for the quasi-deuteron in Nucleus\label{Formalism}}
We formulate the formation cross section of the
heavy meson bound states by the two nucleon pick-up reactions.
As we will see below, our model is so simple that it can be applied 
generally to the heavy meson bound state formations by the
two nucleon pick-up reactions such as ($\gamma, d$) and ($p, ^3$He). 

We apply the effective number approach, which has been used for the
studies of 
the meson-nucleus bound states~\cite{d3He-nd}, 
to evaluate the formation rate of the bound systems in the two nucleon
pick-up reactions.
In the effective number approach, the formation cross section by
the two nucleon pick-up reactions can be written as,
\begin{equation}
\frac{d^2\sigma}{dE d\Omega}=\left(\frac{d\sigma}{d\Omega}\right)^{{\rm ele}}
\sum_{f} \frac{\Gamma}{2\pi}\frac{1}{\Delta E^2 +
{\Gamma^2} /4} N_{{\rm eff}},
\label{Cross}
\end{equation}
where $(d\sigma/d\Omega)^{\rm ele}$ is the elementary cross section of
the meson production, and $\Gamma$ the width of the meson
bound states.
The all combinations of the final states, 
labeled by $f$, are summed
up to evaluate the inclusive cross section.
The energy transfer $\Delta E$ of the reaction in the laboratory
frame is defined as,
\begin{eqnarray}
\Delta E 
=(T_f + M_f) + (M-B) - (T_i + M_i) - (2M_N - S_{2N}), \nonumber\\  
\label{DelE}
\end{eqnarray}
where 
$T_f$ and $M_f$ are the kinetic energy and the mass of the emitted particle,   
$T_i$ and $M_i$ the kinetic energy and the mass of the incident particle, and 
$M$ the mass of the produced meson.
The meson binding energy $B$ and the two nucleon separation
energy $S_{2N}$ from the target nucleus are determined for each bound level of
meson and excited level of the daughter nucleus.
Here, we neglect the recoil energy of the nucleus.

\begin{figure}[tb]
\begin{center}
\includegraphics [width=7.0cm, height=5.0cm]{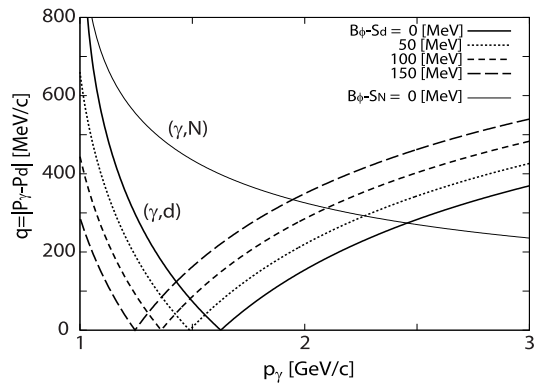}
\caption{Momentum transfer ${\bm q}$ of the forward ($\gamma,d$)
reaction for the formation of $\phi(1020)$ meson bound states in the heavy
target nucleus plotted as functions of the incident photon momentum
$p_{\gamma}$ for four values of the gap ($B_\phi-S_d$) between the $\phi$ meson
binding energy $B_{\phi}$ and the deuteron separation energy $S_{d}$ as
indicated in the figure.
Momentum transfer of the one-nucleon pick-up ($\gamma,N$) reaction is
also shown for comparison. 
\label{q(gamma)}}
\end{center}
\end{figure}

\begin{figure}
\begin{center}
\includegraphics [width=7.0cm, height=5cm]{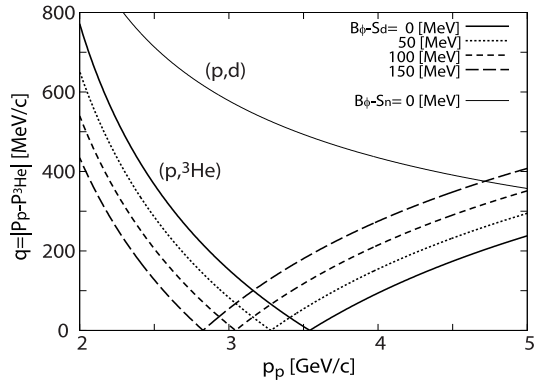}
\caption{Same as Fig.~\ref{q(gamma)} except for the proton induced ($p,^{3}$He) and
 ($p,d$) reactions.
\label{q(p)}}
\end{center}
\end{figure}

\begin{figure}[tb]
\begin{center}
\includegraphics [width=7.0cm, height=5.5cm]{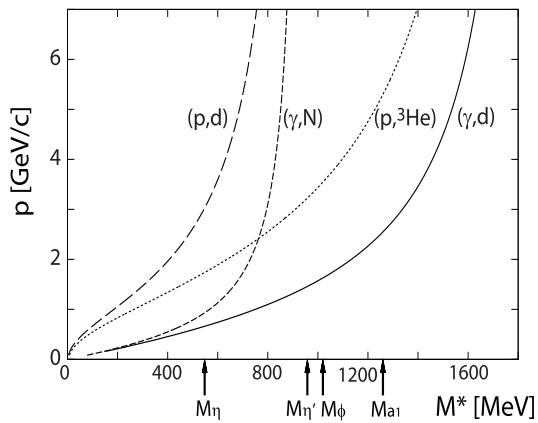}
\caption{Incident beam momenta $p$ required to produce meson in
recoilless kinematics in one nucleon pick-up (($\gamma,N$), ($p,d$)) and
two nucleon pick-up (($\gamma,d$), ($p$,$^3$He)) reactions are plotted as
functions of meson effective mass
$M^{*}$ which is defined as $M^{*}=M-B$ with the meson mass $M$ and the
binding energy $B$.
The in-vacuum masses of $\eta$, $\eta^{\prime}$, $\phi$ and $a_1$ are
indicated in the figure by arrows.
\label{q-line}}
\end{center}
\end{figure}

The momentum transfer ${\bm q}$ of the reaction is defined as,
\begin{equation}
{\bm q}={\bm p}_i-{\bm p}_f,
\end{equation}
and shown in Fig.~\ref{q(gamma)} for the $\phi$ meson
formation case, as an example,
in the heavy target for the ($\gamma,d$) reaction together with
that for one-nucleon pick-up ($\gamma,N$) reaction
as functions of the momentum of the incident photon.
We also show the momentum transfers for the ($p,^3$He) and ($p,d$)
reactions for the $\phi$ meson formation in Fig.~\ref{q(p)} for
comparison.
As we can see from the figures, the two-nucleon pick-up reactions satisfy the
recoilless condition ${\bm q}=0$ at a finite incident momentum, while 
large momentum transfer is 
unavoidable for the one-nucleon pick-up reactions.
We show in Fig.~\ref{q-line} the incident beam momenta $p$ for 
the four reactions required to produce the meson with an effective mass
$M^{*}$ in recoilless kinematics.
The effective mass $M^{*}$ is defined as $M^{*}=M-B$ with the meson mass
$M$ and the binding energy $B$. 
We find clearly that the meson with larger effective mass than nucleon
can not be produced in recoilless kinematics by the one nucleon pick-up
reactions.
In two nucleon  pick-up reactions like ($\gamma,d$) and ($p, ^3$He), on
the other hand, we can produce the heavier meson states like
$\eta^{\prime}(958)$, $\phi(1020)$, and $a_1 (1260)$ meson bound states in
the recoilless kinematics.

The theoretical calculation of the two nucleon pick-up reactions is
rather difficult in general.
In addition, we expect complicated nuclear excitations in daughter
nuclei with two nucleon holes, which will prevent us from the clear
identification of meson bound states.
Here, we consider the specific nucleus $^6$Li as the target which has
the well-developed cluster structure of $\alpha + d$ in the ground
state.
The probability of the $\alpha+d$ component of the ground state of 
$^6$Li is reported to be 0.616 in Ref.~\cite{Parke} and 0.73 in
Ref.~\cite{6Li}.
In the reaction considered in this article, the momentum transfer to the
initial deuteron wavefunction in $^6$Li 
and the final meson wavefunction in the mesic nucleus
is considered to be small near the recoilless kinematics and 
the quasi-deuteron picture in the $^6$Li target is expected to
be a good approximation.
Thus, by considering the nuclei like $^6$Li which have the large component of
quasi-deuteron as targets, we can evaluate the reaction rate in a simple
way and expect to have the simple structure of the formation spectra of
mesic nuclei in the two nucleon pick-up reactions.
In our model considered, we treat $^6$Li as
the bound state of the alpha particle and the deuteron 
to evaluate the reaction rate.
Thus, $B$ in Eq.~(\ref{DelE}) is the binding
energy of a heavy meson and $\alpha$ particle in the final state and
$S_{2N}$ is fixed to be $S_{2N}=1.47$ [MeV] from the mass gap of 
initial and
final nuclei as $S_{2N}=(M_\alpha+M_d)-M_{^6{\rm Li}}$.
The elementary cross section
$(d\sigma/d\Omega)^{\rm ele}$ for the meson production
appearing in Eq.~(\ref{Cross}) is that of the 
$i + d \rightarrow f + {\rm meson}$ reaction 
with the incident particle $i$ and the emitted particle $f$,
which should be evaluated from experiments as previous 
cases~\cite{d3He-nd,K,eta}.
Here, the emitted particle $f$ will be $^3$He for the proton induced
($i=$ proton) case, and $d$ for the $\gamma$ induced 
($i=$ photon) case, respectively.

\begin{figure}[tb]
\begin{center}
\includegraphics [width=8cm, height=5.5cm]{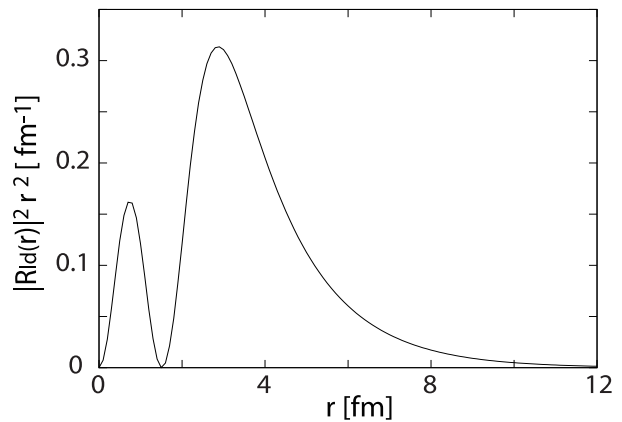}
\caption{Calculated density distribution of the radial part
$R_{l_d}(r)$ of the relative wavefunction $\psi_{l_d}({\bm r})$ of
 $\alpha$ and deuteron in $^6$Li nucleus.
\label{WF(a-d)}}
\end{center}
\end{figure}

\begin{figure}[tb]
\begin{center}
\includegraphics [width=7.6cm, height=10cm]{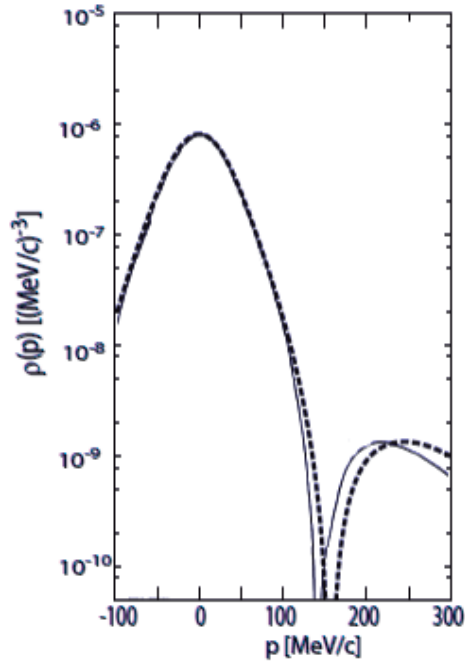}
\caption{Momentum distribution of deuteron in $^6$Li obtained by our
model (dashed line) and by the analysis of the experimental
data~\cite{6Li} (solid line). 
The correction factor 0.73 is multiplied to the results of our model to
satisfy the same normalization as in Ref.~\cite{6Li}.  
\label{Mom(a-d)}}
\end{center}
\end{figure}

In our model considering the $^{6}$Li nucleus as the bound state of alpha
particle and deuteron, the effective number $N_{\rm eff}$ of the
$^{6}$Li($i,f)\alpha \otimes $meson reaction can be written as,  
\begin{equation}
 N_{{\rm eff}}=\sum_{JM} {\mid \int \chi^{\ast}_{f}({\bm r})
  [\phi^{\ast}_{l_m} ({\bm r}) \otimes \psi_{l_{d}}({\bm r})]_{JM}
  \chi_i ({\bm r}) 
d{\bm r} \mid}^2,
\label{Neff} 
\end{equation}
where $\phi_{l_m} ({\bm r})$ is the wavefunction of the
meson bound state and
$\psi_{l_d}({\bm r})$ that of the deuteron bound to $\alpha$
in the $^6$Li target.
$\chi_i ({\bm r})$ and $\chi_f ({\bm r})$ are the
incident and the emitted particle wavefunctions in
the scattering states, respectively.
We assume plane waves for $\chi_i$ and $\chi_f$ in this
article.
The distortion effects to $\chi_i$ and $\chi_f$ depend on the incident
and emitted particles~\cite{gammap}, however they are known to be relatively
small for the cases satisfying the matching condition~\cite{d3He-nd}.
The deuteron wavefunction $\psi_{l_d}$ in $^6$Li is determined to
reproduce the momentum distribution reported in Ref.~\cite{6Li} based on
the analysis of the $^6$Li$(e,e^{\prime}d)^4$He reaction.
We calculate the $\psi_{l_d}$ by solving the Schr\"{o}dinger equation with the
Woods-Saxon type potential,
\begin{equation}
U(r)=\frac{U_{0}}{1+ \exp[(r-R)/a]},
\label{Uopt(a-d)}
\end{equation}
and adjust the potential depth $U_0$ and the radius parameter $R$ to
reproduce the momentum
distribution $\rho(p)$ reported in Ref.~\cite{6Li}.
$\rho(p)$ is defined as,
\begin{equation}
\rho(p)=\frac{1}{(2\pi)^3} \left| \int e^{-i {\bm p} \cdot {\bm r}} \psi_{l_d}({\bm r}) d{\bm r} \right|^2,
\label{dens(a-d)}
\end{equation}
where $|\psi_{l_d}({\bm r})|^2$ is normalized to be 1 in the
coordinate space as usual.
The potential parameters used here are fixed to be $R=2.0$ [fm], $a=0.5$
[fm], and $U_{0}=-75$ [MeV].
The calculated wavefunction is shown in Fig~\ref{WF(a-d)} and momentum
distribution in Fig.~\ref{Mom(a-d)}.
The wavefunction in Fig.~\ref{WF(a-d)} corresponds to the 2$s$ bound
state as indicated in Ref.~\cite{6Li}
because of the Pauli effect to nucleons which forbids the alpha and
deuteron clusters to be in the relative 1$s$ state.
The calculated momentum
distribution reproduce the PWIA result in Ref.~\cite{6Li} reasonably well as
shown in the Fig.~\ref{Mom(a-d)}.
In Fig.~\ref{Mom(a-d)}, we multiply a factor 0.73 to our
results to correct the overall normalization of $\rho(p)$ to be the same as
Ref.~\cite{6Li}.
It should be noted that the PWIA results in Ref.~\cite{6Li} ignoring the
distortion effects are
the quantity which should be compared with our results calculated by
Eq.~(\ref{dens(a-d)}).

\begin{figure}[tb]
\begin{center}
\includegraphics [width=8cm, height=6cm]{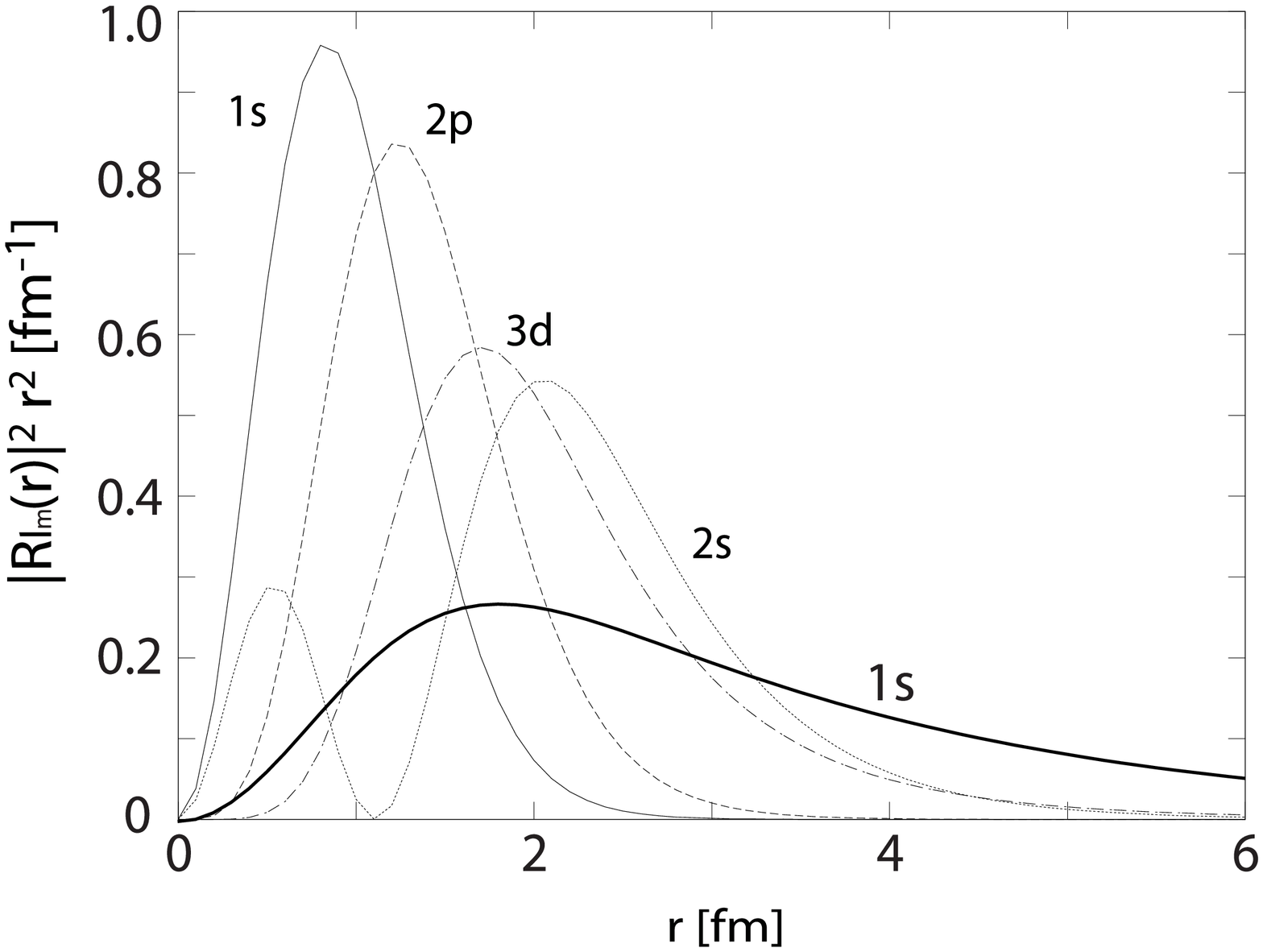}
\caption{Calculated density distribution of the radial part $R_{l_m}(r)$
of the wavefunction $\phi_{l_m}({\bm r})$ of the meson bound
 state in the $\alpha$ particle 
for the potential strength ($V_0, W_0$)=($-34.7, -7.5$) [MeV] (thick solid
line) and ($-250$, $-5$) [MeV] (thin lines).
\label{WF(phi)}}
\end{center}
\end{figure}

The bound meson wavefunctions $\phi_{l_m}({\bm r})$ in the final state are calculated
by solving the Klein-Gordon equation with the optical potential in the
Woods-Saxon form as in Eq.~(\ref{Uopt(a-d)}).
We fix the meson mass to be $M=1020$ [MeV] as $\phi$ meson, which is
heavier than nucleon and cannot be formed in the recoilless kinematics in one
nucleon pick-up reactions.
Since mesons can be absorbed by the nucleus generally,
the potential
strength $U_0$ in Eq.~(\ref{Uopt(a-d)}) is considered to be a complex
number as $U_0=(V_0+iW_0$) for mesons. 
In the present calculation, 
we consider a few potential strengths as examples and study how the two
nucleon pick-up reaction spectra change according to the meson-nucleus 
interaction.
We consider first the potential strengths based on the experimental data
obtained by E325 in KEK~\cite{Muto}, where the $\phi$ meson mass shift
is reported as $\Delta m(\rho_0)/m=-3.4$ \% and the $\phi$ meson width
in nucleus $\Gamma_\phi (\rho_0)=15$ [MeV].
We adopt these numbers as the $\phi$ mesic optical potential and fix as
$(V_0, W_0) =(-34.7, -7.5)$ [MeV].
We mention here that the large in-medium width $\Gamma_{\phi}(\rho_0)
\simeq 80$ [MeV] of $\phi$ meson was also indicated based on 
another attenuation
experiment by LEPS at SPring-8~\cite{Isikawa}, which corresponds to the
imaginary potential strength $W_0=-40$ [MeV].
We have checked numerically that the potential with this imaginary
strength together with the real part strength corresponding to the 3.4
\% mass reduction does not provide $\phi$ meson bound states in alpha
potential.
We, then, assume a stronger attractive potential for meson-nucleus system
to estimate the formation rate of heavy meson bound systems with strong
attractive potential as reported in Ref.~\cite{Nagahiro2} for 
$\eta^{\prime}(958)$ meson in a theoretical model. 
The assumed potential parameters with two different absorption strength
are $(V_0, W_0) = (-250, -5)$ and ($-250, -20$) in unit of MeV.
The distribution parameters are fixed to be $R=1.18A^{1/3}-0.48$ [fm] and
$a=0.5$ [fm] with $A=4$ for meson-alpha system. 
We show in Table~\ref{BEphi} the calculated binding energies and widths
of the meson bound states for the three different potentials.
The radial density distributions are also shown in Fig.~\ref{WF(phi)}
for  $(V_0, W_0) =(-34.7, -7.5)$ and ($-250, -5$) [MeV] cases.
We found the radial density distributions of 1$s$ states of the both
potentials are much different because of the different potential
strength.

%\begin{table*}
\begin{table}[tb]
\caption{Calculated binding energies and widths of the 
 meson$-$alpha bound states in unit of MeV with the potential strength ($V_0,W_0$) = 
(i) ($-34.7, -7.5$), (ii) ($-250$, $-5$), and (iii) ($-250$, $-20$) [MeV].
The meson mass is fixed to be 1020 MeV.
\label{BEphi}}
\begin{center}
\begin{ruledtabular}
\begin{tabular}{c|cc|cc|cc} %\hline\hline
\multicolumn{1}{c|}{($n_{\phi}, \ell_{\phi})$}%
&\multicolumn{2}{c|} {(i)} 
&\multicolumn{2}{c|} {(ii)} 
&\multicolumn{2}{c} {(iii)}  \\ 
\multicolumn{1}{c|}{state} & B.E. & $\Gamma$  &
 B.E.  & $\Gamma$  & B.E.  & $\Gamma$   \\ \hline
1$s$& 0.76 & 2.9  & 131.5 & 8.1 & 131.2 & 32.5 \\
2$s$&      &      &  14.5 & 2.8 &  14.2 & 11.2 \\ 
2$p$&      &      &  58.0 & 5.5 &  57.7 & 21.8 \\
3$d$&      &      &   2.9 & 2.8 &  2.4  & 11.3 \\ %\hline\hline	   
\end{tabular}
\end{ruledtabular}
%\end{tabular}
\end{center}
%\end{table*} 
\end{table} 

Since the angular momentum $l_d$ of the relative wavefunction of deuteron and
$\alpha$ particle in $^6$Li is considered to be 0, the expression of the effective numbers in
Eq.~(\ref{Neff}) can be simplified as, 
\begin{equation}
 N_{{\rm eff}}=\sum_{M} {\mid \int e^{i{\bm q}\cdot
  \boldsymbol{r}}\phi^{\ast}_{l_m} ({\bm r}) \psi_0({\bm r}) d \boldsymbol{r} \mid}^2,
\label{Neff2}
\end{equation} 
in the plane wave approximation.
We use this expression to calculation the effective numbers in this article.

In the recoilless kinematics, the only $s$ states of the meson bound
states can be populated because of the orthogonality of the angular
part of the wavefunction as can be seen in Eq.~(\ref{Neff2}).
And because of the approximate orthogonality of the radial parts of
$\phi_{l_m}$ and $\psi_0$, the
substitutional 2$s$ state of the meson is expected be largely
populated.

\section{Numerical Result for Bound state Formation rate \label{Result}}
\begin{figure}[tb]
\begin{center}
\includegraphics [width=8cm, height=6cm]{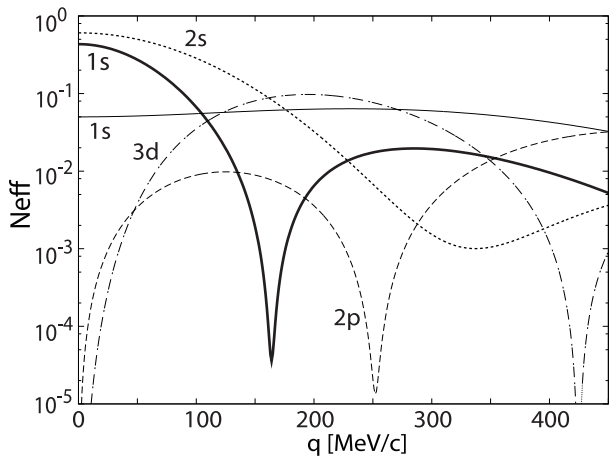}
\caption{Calculated effective numbers are plotted as functions of the
momentum transfer of the two nucleon pick-up reactions for $^6$Li target
for the 1$s$ meson bound state formation with the potential strength
($V_0, W_0$) = ($-34.7, -7.5$) [MeV] (thick solid line) and
the 1$s$, 2$s$, 2$p$, and 3$d$ bound states 
with ($V_0, W_0$) = ($-250, -5$) [MeV] (thin lines).
\label{Neff-q}}
\end{center}
\end{figure}

%subsection{}
%subsubsection{}
To investigate the momentum transfer dependence of $N_{\rm eff}$ for
each bound state formation, 
we show in Fig.~\ref{Neff-q} the calculated effective numbers $N_{\rm
eff}$ by Eq.~(\ref{Neff2}) as functions of the momentum transfer 
$|{\bm q}|$ for two nucleon pick-up reactions for $^6$Li target.
Each effective number has the characteristic behavior due to the
matching condition of the momentum transfer and the angular momentum
transfer.
As we have mentioned in the previous section, the effective numbers 
$N_{\rm eff}$ is
exactly 0 for 2$p$ and 3$d$ states at $|{\bm q}|=0$ because of the
orthogonality condition of the angular part wavefunction to the $s$-wave
function $\psi_0$.
And the substitutional 2$s$ state of the bound meson have the largest
contribution at $|{\bm q}|=0$ as we expected.
The contribution of the 1$s$ state with weaker real potential $V_0 =-34.7$
[MeV] has the stronger dependence on $q$ as naturally expected its larger
special dimensions as shown in Fig.~\ref{WF(phi)}.
As the momentum transfer increases, the 3$d$ bound state formation has the largest
contribution at $160 \lesssim |{\bm q}| \lesssim 270$ [MeV/c] and, then
the $1s$ bound state at $270 \lesssim |{\bm q}| \lesssim 450$ [MeV/c].
Overall strength of the heavy meson bound state formation cross section
becomes smaller for the kinematics with larger momentum transfer.

\begin{figure}
\begin{center}
\includegraphics [width=8cm, height=6.0cm]{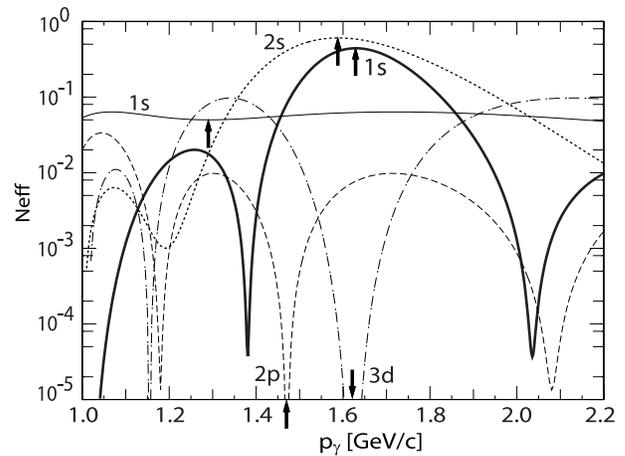}
\caption{Same as Fig.~\ref{Neff-q} except for the plots as functions of
the incident photon momentum $p_\gamma$ for ($\gamma,d$) reaction.
The arrow indicates the incident photon momentum of the recoilless kinematics for
each meson bound state formation.
\label{Neff-Pg}}
\end{center}
\end{figure}

\begin{figure}
\begin{center}
\includegraphics [width=8cm, height=6.0cm]{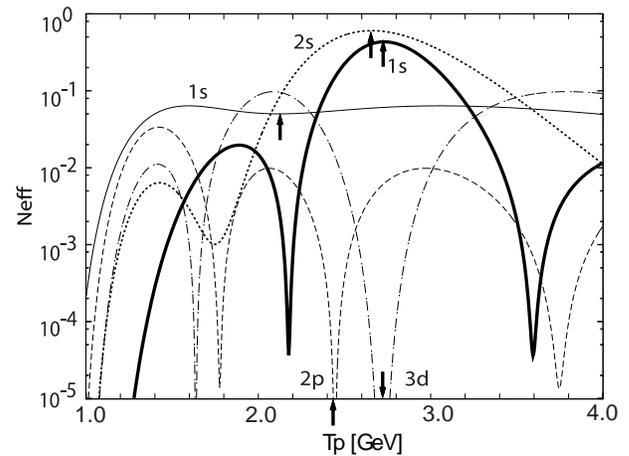}
\caption{
Same as Fig.~\ref{Neff-q} except for the plots as functions of
the incident proton kinetic energy $T_p$ for ($p,{{^3}{\rm He}}$) reaction.
The arrow indicates the incident proton kinetic energy of the recoilless
kinematics for each meson bound state formation.
\label{Neff-Tp}}
\end{center}
\end{figure}

\begin{figure*}
\begin{center}
\includegraphics [width=16.5cm, height=12cm]{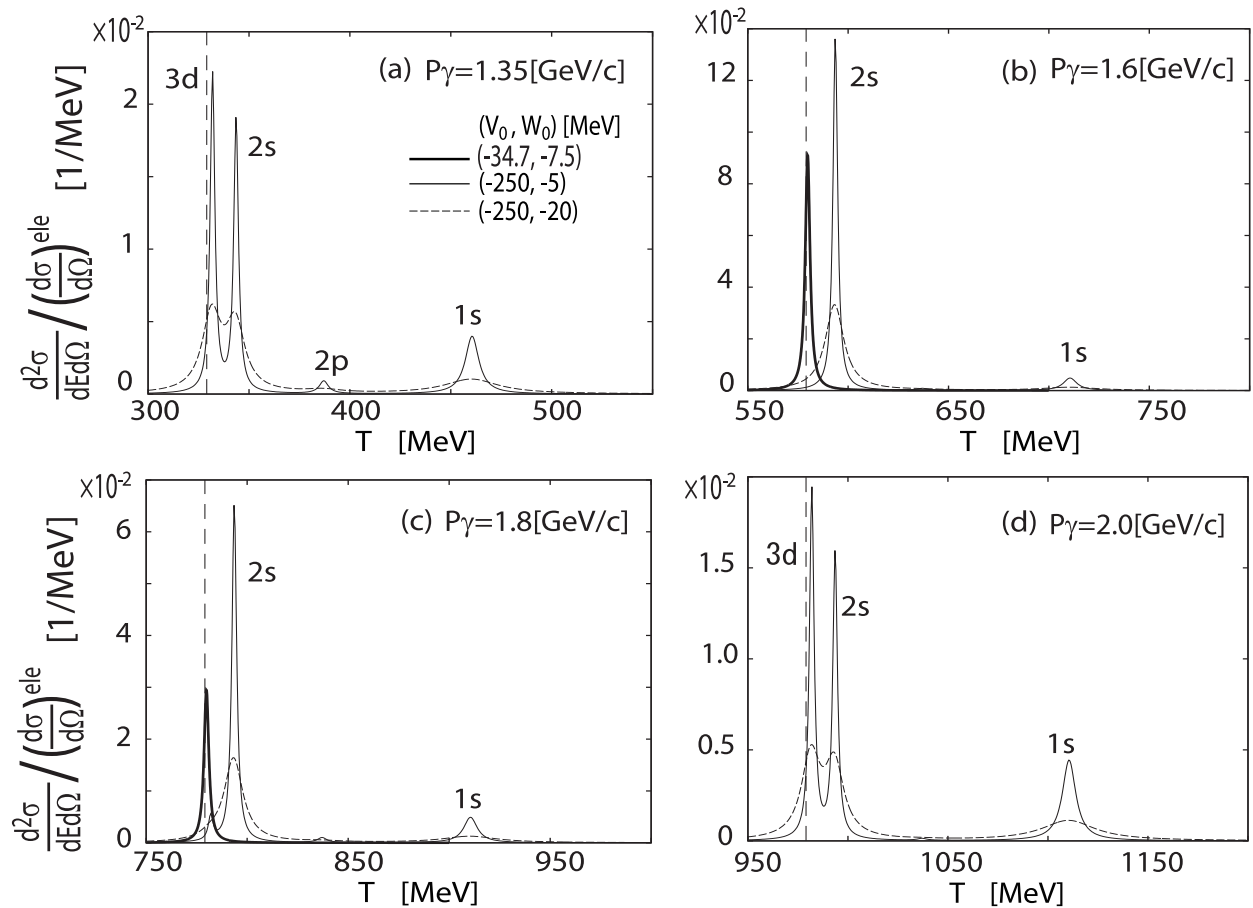}
\caption{
Expected spectra of the forward two nucleon pick-up ($\gamma,d$) reaction 
$^6$Li for the
formation of the meson$-\alpha$ bound states are plotted as functions of
the emitted deuteron kinetic energy for the incident photon momenta
$p_{\gamma}=$ $(a)$ $1.35$, $(b)$ $1.6$, $(c)$ $1.8$, and $(d)$ $2.0$ [GeV/c].
The vertical dashed line indicates the meson production
threshold.
Each line is calculated with the different optical potential parameters 
for meson$-\alpha$ system as indicated in $(a)$.
The spectrum with the potential $(V_0, W_0)=(-34.7, 7.5)$ [MeV] is only
shown in (b) and (c) because it is small and invisible in (a) and (d). 
\label{Cross(phi)}}
\end{center}
\end{figure*}

We show the same effective numbers as functions of the incident particle
energies for ($\gamma,d$) and ($p,{^3{\rm He}}$) reaction cases 
in Figs.~\ref{Neff-Pg} and \ref{Neff-Tp}.
Because of the different binding energies of the meson appearing in
Eq.~(\ref{DelE}), the incident particle energy which corresponds to recoilless
kinematics ($|{\bm q}|=0$) is different for each bound state.
We indicate the incident particle energy of the recoilless kinematics
for each state by the arrow in Figs.~\ref{Neff-Pg} and \ref{Neff-Tp}.
The both figures show the very similar behavior of $N_{\rm eff}$.

We find that the substitutional 2$s$ state of the meson is
produced in the almost recoilless condition and has the largest
contribution to the cross section at $p_{\gamma}=1.6$ [GeV/c]
($T_p=2.7$ [GeV]).
(In the explanation below, we indicate the photon momentum $p_\gamma$ in
($\gamma,d$) reaction with the corresponding proton kinetic energy $T_p$ 
in the ($p,^3$He) reaction in the parenthesis.)
We mention here that the 1$s$ state wavefunction with 
the potential strength $(V_0, W_0)=(-34.7, -7.5)$ [MeV]
has larger spatial dimension which violates the approximate
orthogonality of the radial part with $\psi_0$ at $q=0$ in
Eq.~(\ref{Neff2}), 
and thus, the contribution of this 1$s$ state also has
the large contribution at $p_{\gamma}=1.6$ [GeV/c] ($T_p=2.7$ [GeV]).
At $p_{\gamma}=1.35$ [GeV/c] ($T_p=2.1$ [GeV]), a little above the 
$\phi$ production
threshold of the elementary process, the size of the
effective number of
the 2$s$ and 3$d$ states are similar and larger than those of other
states formation.
At $p_{\gamma}=1.8$ [GeV/c] ($T_p=3.1 $ [GeV]), 
the effective number of the 2$s$ state
formation is still dominant, however the other contributions of the 1$s$
and 3$d$ states formation become relatively more important than at
$p_{\gamma}=1.6$ [GeV/c] ($T_p=2.7$ [GeV]).
The contributions of the 1$s$, 2$s$, and 3$d$ state formations are
important at $p_{\gamma}=2.0$ [GeV/c] ($T_p=3.5$ [GeV]) and the 
formation spectrum is
expected to be a little more complicated than those at other photon
momenta.
We find that $N_{\rm eff}$ for 2$s$ bound state formation 
takes the largest value $ N_{\rm eff}=0.606$
at $p_{\gamma}=1.59$ [GeV/c] ($T_p =2.66$ [GeV]) which means that the
meson bound state formation cross section can be about half of the
elementary cross section.

We then calculate the relative strength of the formation spectra of 
the meson bound states
in alpha particle by the two nucleon pick-up reaction in
$^6$Li target at incident
photon momenta $p_\gamma=1.35, 1.6, 1.8$ and 2.0 [GeV/c] for
the ($\gamma,d$) reaction which corresponds to the 
$T_p =2.1, 2.7, 3.1$ and $3.5$ [GeV] for
the ($p,{{^3}{\rm He}}$) reaction.
We can expect to observe the different behavior of the formation spectra
at these energies as expected from the energy dependence of the
effective numbers.
The calculated results 
$\displaystyle  \frac{d^2\sigma}{dEd\Omega} 
{\Big /} \left ( \frac{d\sigma}{d\Omega} \right)^{\rm ele}$
for ($\gamma,d$) reaction are shown in 
Fig.~\ref{Cross(phi)} for three
different optical potential parameters for the meson bound in $\alpha$
particle.

The expected spectra for the potential strength 
$(V_0, W_0)=(-34.7, -7.5)$ [MeV] case are simple since there is only 
lightly bound 1$s$ state.
As shown in Fig.~\ref{Cross(phi)} (b) and (c), this 1$s$ state is seen as a
peak close to this meson production threshold for $p_{\gamma}=1.6$ and
1.8 [GeV/c] ($T_p=2.7$ and 3.1 [GeV]).
The contribution of this state has so strong $q$ and incident energy
dependence as shown in Figs.~\ref{Neff-q}-\ref{Neff-Tp} and
that it becomes
smaller than those with deeper potential cases and
invisible in Fig.~\ref{Cross(phi)} (a) and (d).

For deeper potential cases with $(V_0, W_0)=(-250, -5)$ and
($-250, -20$) [MeV],
we find that the spectra at $p_{\gamma}=1.6$ and 1.8 [GeV/c]
($T_p=2.7$ and $3.1$ [GeV]) are
dominated by the 2$s$ state formation and the other contributions are
significantly small.
On the other hand, we can observe clear peak structures of the 1$s$ and
3$d$ states formation in addition to the 2$s$ state at $p_{\gamma}=1.35$
and 2.0 [GeV/c] ($T_p=2.1$ and $3.5$ [GeV]).
The size of the spectra 
$\displaystyle  \frac{d^2\sigma}{dEd\Omega} 
{\Big /} \left ( \frac{d\sigma}{d\Omega} \right)^{\rm ele}$ 
are relatively large at
$p_{\gamma}=1.6$ and 1.8 [GeV/c]  ($T_p=2.7$ and $3.1$ [GeV])
where the momentum transfer of the
two nucleon pick-up reactions is small.
At $p_{\gamma}=1.8$ and 2.0 [GeV/c] ($T_p=3.1$ and $3.5$ [GeV]), 
the momentum transfer is larger and
the size of the spectra become smaller rapidly for the
larger incident momentum and energy.
We also show the effects of the imaginary part of the optical potential
to the formation spectra in Fig.~\ref{Cross(phi)}.
Since we have only one deuteron state in the initial nucleus $^6$Li, the
reaction spectra have simple structure, especially for 
$(V_0, W_0)=(-34.7, -7.5)$ [MeV] potential case.
Thus, the main effects of the absorptive potential are found to reduce
the peak height with large width.
The overlap of the resonance peaks due to the widths only happens
between 2$s$ and 3$d$ states for the absorption potential strengths
studied here in the two nucleon pick-up spectra. 

In case if the imaginary potential is large as $W_0=-40$ [MeV] as
indicated based on the data reported for the $\phi$ meson 
in Ref.~\cite{Isikawa}, 
the height of all peaks in the spectra shown in Fig.~\ref{Cross(phi)}
becomes lower
in inverse proportion to the strength of the imaginary potential and
the contributions of 2$s$ and 3$d$ states for $V_0=-250$ [MeV] case
can not be distinguished because of the large widths.

\section{Summary}
We have considered the two nucleon pick-up reactions in this
article to investigate the feasibility of the reactions of this type to
extend our research field of the meson nucleus bound systems to the
heavier meson region.
We have developed a model and used the effective number approach to
evaluate the formation rate.
As an example, we have applied the model to the heavy meson bound state
formation in alpha particle with $^6$Li target.
As shown in the numerical results, we have found that the shape of the
formation spectra is simple and seems to be suited to extract the
binding energies and widths of the meson bound state because of the
simple $\alpha-d$ cluster structure of $^6$Li.
The size of the formation cross section can be more than half
of the elementary cross section at the recoilless kinematics.

This theoretical model is so simple that we can apply it easily to
evaluate the mesic nucleus formation rate of other two nucleon pick-up
reactions such as ($\pi,d$) and so on.
To do this,
we should simply replace the elementary cross sections to those of the
appropriate processes of the meson production like 
$\pi + d \rightarrow d + $ heavy meson.
In this model, however, the target nucleus is required to have large
component of
the quasi deuteron structure.
Thus, we have considered the $^6$Li target as an example in this article.
The calculated spectra shape are expected to have simple structure
generally because of the existence of the quasi-deuteron in target
nucleus and are suited to extract meson properties from the reaction
spectra.

In general cases, we should not assume the existence of the quasi
deuteron in the target nuclei~\cite{Simpson} and 
we need to evaluate the emissions of the deuteron composed of two
nucleons which are in the different single particle levels in the target.
The deuteron can be formed from any pairs of proton and neutron in the target 
by the reaction of the meson production.
In this process, however, the spectrum shape could be so complicated 
that it is difficult to extract meson properties.

As for the actual experimental observations, the calculated cross
sections could be too small to find peak structures in the inclusive missing
mass spectra due to the size of the elementary cross section, 
and the coincident measurement detecting the particle pair
emissions from meson absorption in nucleus may be necessary to reduce
the background.
So far, the formation $\eta$ mesic nucleus in the two nucleon pick-up
($p, ^3$He) reaction for $^{27}$Al target was reported in
Ref~\cite{Budzanowski} by COSY-GEM.
They performed the coincidence measurement with $p \pi^-$ pair
emissions from $\eta N$ in nucleus.
They reported 0.5 [nb] for the upper limit of the signal of the $\eta$
mesic nuclear formation cross section~\cite{Budzanowski} at the energy
where the elementary cross section is 77 [nb/sr]~\cite{Berthet}.
In our example considered in this article, 
the effective number $N_{\rm eff}$ for the formation of meson bound
state is about 0.5 for both shallow and deep potential cases around
$p_{\gamma}=1.6$ [GeV/c] ($T_p =2.7$ [GeV]) for the largest
contributions as shown in Fig.~\ref{Neff-Pg}.
In this sense, we think the present result also have relevance as
a guide for the actual experiment.

We believe that it is quite important to find new reactions suited 
to form the heavy meson bound states in nucleus to explore the
various aspects of the strong interaction symmetries at finite density
by the mesic nuclei.
In this context, the two nucleon pick-up reactions studied in this
article are quite interesting, 
since we can satisfy the recoilless condition in this reaction for the
formation of meson heavier than nucleon.

\begin{acknowledgments}
We acknowledge the fruitful discussions with H. Fujioka and K. Itahashi.
N. I. appreciates the support by the Grant-in-Aid for JSPS Fellows.
This work was partly supported by the Grants-in-Aid for Scientific Research 
(No.~22740161, No.~20540273, and No.~22105510). 
This work was done in part under the 
Yukawa International Program for Quark-hadron Sciences (YIPQS).
\end{acknowledgments}


\begin{thebibliography}{99}\label{sec:TeXbooks}
\bibitem{QCD}T. Hatsuda and T. Kunihiro, Phys. Rept. \textbf{247}, 221 (1994), and references therein.
\bibitem{Jido}D. Jido, T. Hatsuda and T. Kunihiro, Phys. Lett. {\bf B670}, 109 (2008).
\bibitem{Kolo}E. E. Kolomeitsev, N. Kaiser, and W. Weise, Phys. Rev. Lett. {\bf 90}, 092501 (2003).
\bibitem{d3He-nd} H. Toki, S. Hirenzaki, T. Yamazaki and R. S. Hayano, Nucl. Phys. \textbf{A501} (1989), 653;\\ 
S. Hirenzaki, H. Toki, and T. Yamazaki, Phys. Rev. {\bf C44}, 2472 (1991);\\
H. Toki, S. Hirenzaki, and T. Yamazaki, Nucl. Phys. {\bf A530}, 679 (1991).
\bibitem{K}
T. Kishimoto, Phys. Rev. Lett. {\bf 83}, 4701 (1999);\\
S. Hirenzaki, Y. Okumura, H. Toki, E. Oset, and A. Ramos, Phys. Rev. \textbf{C61}, 055205 (2000);\\
K. Ikuta, M. Arima, and K. Masutani, Prog. Theor. Phys. {\bf 108}, 917 (2002);\\ 
J. Yamagata, H. Nagahiro, Y. Okumura, and S. Hirenzaki,	Prog. Theor. Phys. {\bf 114}, 301 (2005) 
[Errata-{\it ibid} {\bf	114}, 905 (2005)];\\
J. Yamagata, H. Nagahiro, and S. Hirenzaki, Phys. Rev. {\bf C74}, 014604 (2006).
\bibitem{eta} 
Q. Haider and L. C. Liu, Phys. Lett. {\bf B172}, 257 (1986);\\
R. S. Hayano, S. Hirenzaki, and A. Gillitzer, Eur. Phys. J. A {\bf 6}, 99 (1999);\\ 
D. Jido, H. Nagahiro, and S. Hirenzaki, Phys. Rev. {\bf C66}, 045202 (2002);\\
H. Nagahiro, D. Jido, and S. Hirenzaki, Phys. Rev. {\bf C68}, 035205 (2003);\\
D. Jido, E. E. Kolomeitsev, H. Nagahiro, and S. Hirenzaki, Nucl. Phys. {\bf A811}, 158 (2008).
\bibitem{K.Suzuki} K. Suzuki  {\it et al.}, Phys. Rev. Lett. {\bf 92}, 072302 (2004). 
\bibitem{gammap} H. Nagahiro, D. Jido, and S. Hirenzaki, Nucl. Phys. {\bf A761}, 92 (2005).
\bibitem{piN} H. Nagahiro, D. Jido, and S. Hirenzaki, Phys. Rev. {\bf C80}, 025205 (2009).
\bibitem{Nagahiro} H. Nagahiro and S. Hirenzaki, Phys. Rev. Lett. {\bf 94}, 232503 (2005).
\bibitem{Nagahiro2}H. Nagahiro, M. Takizawa, and S. Hirenzaki, Phys. Rev. {\bf C74}, 045203 (2006).
\bibitem{Nagahiro3}D. Jido, H. Nagahiro and S. Hirenzaki, [arXiv:1109.0394 [nucl-th]].

\bibitem{Yamagata} J. Yamagata-Sekihara, S. Hirenzaki, D. Cabrera, and M. J. Vicente-Vacas, Prog. Theor. Phys. {\bf 124}, 147 (2010). 
\bibitem{Garcia} C. Garc\'ia-Recio, J. Nieves, and L. Tolos, Phys. Lett. {\bf B690}, 369 (2010).
\bibitem{Budzanowski} A. Budzanowski {\it et al.}, Phys. Rev. {\bf C79}, 012201(R) (2009).
\bibitem{Parke} W. C. Parke and D. R. Lehman, Phys. Rev. {\bf C29}, 2319 (1984).
\bibitem{6Li} R. Ent {\it et al.}, Phys. Rev. Lett. {\bf 57}, 2367 (1986).
\bibitem{Muto} R. Muto {\it et al.} [KEK-PS E325 Collaboration], Phys. Rev. Lett.  {\bf 98}, 042501 (2007).
\bibitem{Isikawa} T. Ishikawa  {\it et al.}, Phys. Lett. {\bf B608}, 215 (2005).
%\bibitem{LEPS} W. C. Chang {\it et al.}, LEPS Collaboration, Phys. Lett. {\bf B658}, 209 (2008).
%\bibitem{CLAS} T. Mibe {\it et al.}, CLAS Collaboration, Phys. Rev. {\bf C76}, 052202(R) (2007).
\bibitem{Simpson} E. C. Simpson and J. A. Tostevin, Phys. Rev. {\bf C83}, 014605 (2011).
\bibitem{Berthet} P. Berthet {\it et al.}, Nucl. Phys. {\bf A443}, 589 (1985).
%\bibitem{Bonn} R. Machileidt {\it et al.}, Phys. Rep. {\bf 149}, 1 (1987).
%\bibitem{pid1} M. Fellinger {\it et al.}, Phys. Rev. Lett. {\bf 22}, 1265 (1969).
%\bibitem{pid2} B. M. Abramov {\it et al.}, Phys. Rep. Lett. {\bf B189}, 295 (1987).
%\bibitem{pid3} B. M. Abramov {\it et al.}, Nucl. Phys. {\bf A542}, 579 (1992).
\end{thebibliography}

\end{document}